\documentclass[11pt]{article}

\usepackage{PRIMEarxiv}
\usepackage[utf8]{inputenc}
\usepackage[T1]{fontenc}
\usepackage{hyperref}
\usepackage{url}
\usepackage{booktabs}
\usepackage{amsfonts}
\usepackage{nicefrac}
\usepackage{microtype}
\usepackage{fancyhdr}
\usepackage{graphicx}
\usepackage{amsmath, amssymb}
\usepackage{subcaption}
\usepackage{float}
\usepackage{adjustbox}
\usepackage[authoryear]{natbib}
\usepackage{rotating}
\usepackage{caption}
\usepackage{multirow} 
\graphicspath{{media/}}

\title{Enhanced NIRMAL Optimizer with Damped Nesterov Acceleration: A Comparative Analysis}

\author{
  Nirmal Gaud \\
  CEO and Master Trainer \\
  ThinkAI - A Machine Learning Community \\
  \texttt{nirmal.gaud.ai@gmail.com} \\
  \And
  Prasad Krishna Murthy \\
  \texttt{prasad.24phddi07@iiitdwd.ac.in} \\
  \And
  Dr. Mostaque Md. Morshedur Hassan \\
  \texttt{mostaq786@gmail.com} \\
  \And
  Abhijit Ganguly \\
  \texttt{abhijitganguly1994@gmail.com} \\
  \And
  Vinay Mali \\
  \texttt{malivinay01234@gmail.com} \\
  \And
  Ms Lalita Bhagwat Randive \\
  \texttt{lalita.bornare@gmail.com} \\
  \And
  Abhaypratap Singh \\
  \texttt{abhayp111@gmail.com} \\
}

\begin{document}
\maketitle

\begin{abstract}
This study introduces the Enhanced NIRMAL (Novel Integrated Robust Multi-Adaptation Learning with Damped Nesterov Acceleration) optimizer, an improved version of the original NIRMAL optimizer. By incorporating an (\(\alpha, r\))-damped Nesterov acceleration mechanism, Enhanced NIRMAL improves convergence stability while retaining chess-inspired strategies of gradient descent, momentum, stochastic perturbations, adaptive learning rates, and non-linear transformations. We evaluate Enhanced NIRMAL against Adam, SGD with Momentum, Nesterov, and the original NIRMAL on four benchmark image classification datasets: MNIST, FashionMNIST, CIFAR-10, and CIFAR-100, using tailored convolutional neural network (CNN) architectures. Enhanced NIRMAL achieves a test accuracy of 46.06\% and the lowest test loss (1.960435) on CIFAR-100, surpassing the original NIRMAL (44.34\% accuracy) and closely rivaling SGD with Momentum (46.43\% accuracy). These results underscore Enhanced NIRMAL's superior generalization and stability, particularly on complex datasets.
\end{abstract}

\keywords{Enhanced NIRMAL \and NIRMAL \and Adam \and SGD with Momentum \and Nesterov \and Optimization \and Deep Learning}

\section{Introduction}
Deep learning models rely on efficient optimization algorithms to minimize losses and improve generalization. This becomes more important when performing complex tasks, such as image classification, and segmentation. Although Stochastic Gradient Descent (SGD) with Momentum and Adam are widely used fundamental optimizers in the field, they can present challenges such as slow convergence rates or high sensitivity to hyperparameter tuning \citep{kingma2017adam, liu2020sgd, acharya2015sgd}. The original NIRMAL (Novel Integrated Robust Multi-Adaptation Learning) optimizer, introduced in \citep{gaud2025nirmal}, proposed a hybrid approach inspired by chess piece movements, integrating gradient descent, momentum, stochastic perturbations, adaptive learning rates, and non-linear transformations. Despite its competitive performance, the convergence stability of NIRMAL in complex datasets needs improvement.

This paper presents Enhanced NIRMAL, incorporating an (\(\alpha, r\))-damped Nesterov acceleration mechanism \citep{cheng2025nesterov} to enhance stability and generalization. We compare Enhanced NIRMAL against Adam, SGD with Momentum, Nesterov, and the original NIRMAL on four datasets: MNIST, FashionMNIST, CIFAR-10, and CIFAR-100. These datasets which vary in complexity, enable a comprehensive evaluation. Custom CNNs were designed, with performance assessed by test accuracy, test loss, and weighted F1-score.

The primary objectives are:
\begin{itemize}
    \item Detail the formulation and enhancements of the Enhanced NIRMAL optimizer.
    \item Evaluate its performance against Adam, SGD with Momentum, Nesterov, and the original NIRMAL on standard image classification benchmarks.
    \item Analyze Enhanced NIRMAL's convergence behavior, stability, and generalization across varying dataset complexities.
\end{itemize}
This work aims to contribute to the development of more robust, efficient, and effective optimization algorithms in deep learning techniques.

\section{Literature Review}

Stochastic gradient-based optimization methods are widely used in machine learning and deep learning to train models efficiently on large datasets. The study in \citep{acharya2015sgd} compares different optimization techniques such as SGD, mini-batch gradient descent, and adaptive optimization using large least squares problems. It was found that mini-batch gradient descent provides a good balance between the noisy updates of standard SGD and the stable updates of full-batch gradient descent. The study also observed that Adam performed best among adaptive optimizers, while Nesterov gave the best results among fixed learning rate methods. It suggested that adaptive optimizers work better on large datasets but may require more computation.

Another study \citep{kingma2017adam} introduced Adam, a popular optimizer that combines ideas from AdaGrad and RMSProp. Adam adapts the learning rate for each parameter by computing exponentially decaying averages of past gradients (first moment) and squared gradients (second moment) \citep{saud2023bert}. It is easy to use, memory-efficient, and works well on large-scale and high-dimensional problems. The authors provided both theoretical analysis and experimental results showing that Adam works well on noisy and non-stationary objectives. They also proposed a variant called AdaMax, which uses a different type of normalization.

The work in \citep{liu2020sgd} focuses on SGD with Momentum (SGDM). Although SGDM is commonly used in practice, its theoretical understanding was limited. This study shows that SGDM can achieve the same convergence rate as SGD in both strongly convex and non-convex settings. It also introduced multistage SGDM, where learning rates and momentum values are changed in stages during training. The authors proved that this approach improves performance and gave theoretical support for a widely used training strategy. They also suggested further research to see whether SGDM can outperform SGD in specific cases.

\citet{botev2016rud} presented Regularized Update Descent (RUD), a framework that derives Nesterov’s Accelerated Gradient (NAG) and classical momentum as approximations to optimize the update direction directly. By minimizing a regularized objective that combines loss and a penalty on the magnitude of the update, RUD provides an intuitive interpretation of NAG’s lookahead mechanism and demonstrates faster convergence than NAG and momentum for quadratic objectives, informing the damping strategy in Enhanced NIRMAL.

\section{Optimization Algorithms}

This section provides a detailed description of the five optimization algorithms evaluated in this study: Enhanced NIRMAL, NIRMAL, Adam, SGD with Momentum and Nestrov. Their mathematical formulations are presented to highlight their unique update mechanisms.

\subsection{Enhanced NIRMAL Optimizer}

NIRMAL's innovative design integrates five distinct strategies, each symbolically associated with a chess piece, contributing a specific aspect to the parameter update:
\begin{itemize}
    \item \textbf{Wazir (Gradient Descent)}: Represents the direct application of the gradient for immediate parameter updates.
    \item \textbf{Elephant (Momentum)}: Incorporates a velocity term to accelerate convergence in consistent directions, smoothing oscillations.
    \item \textbf{Knight (Stochastic Perturbations)}: Introduces random noise to help escape shallow local minima and explore the loss landscape more effectively.
    \item \textbf{Camel (Adaptive Learning Rate Scaling)}: Dynamically adjusts the learning rate based on historical gradient magnitudes, ensuring robust step sizes.
    \item \textbf{Horse (Non-linear Momentum Transformation)}: Applies a non-linear transformation to the momentum term, potentially enhancing stability and controlling step magnitudes.
\end{itemize}

\textbf{Enhanced NIRMAL Optimizer} builds on NIRMAL with an (\(\alpha, r\))-damped Nesterov acceleration mechanism. A damping factor \(\xi_t = r / t^\alpha\) improves stability. The update rules are:
\begin{align}
m_t &= \mu m_{t-1} + (1 - \mu) g_t - \xi_t m_{t-1}, \\
v_t &= \beta v_{t-1} + (1 - \beta) g_t^2, \\
\Delta_{\text{wazir}} &= -\eta g_t, \\
\Delta_{\text{elephant}} &= -\eta m_t, \\
\Delta_{\text{knight}} &= \eta \kappa \mathcal{N}(0, 1), \\
\Delta_{\text{camel}} &= -\eta \gamma \frac{m_t}{\sqrt{v_t + \epsilon}}, \\
\Delta_{\text{horse}} &= -\eta \lambda \tanh(m_t), \\
\Delta_{\text{total}} &= w_{\text{wazir}} \Delta_{\text{wazir}} + w_{\text{elephant}} \Delta_{\text{elephant}} + w_{\text{knight}} \Delta_{\text{knight}} + w_{\text{camel}} \Delta_{\text{camel}} + w_{\text{horse}} \Delta_{\text{horse}}, \\
\theta_{t+1} &= \theta_t + \Delta_{\text{total}}.
\end{align}
Hyperparameters: \(\alpha = 0.5\), \(r = 2.0\), \(\eta = 10^{-3}\), \(\mu = 0.9\), \(\beta = 0.999\), \(\epsilon = 10^{-8}\), \(\kappa = 0.01\), \(\gamma = 1.5\), \(\lambda = 0.5\), weights: \(w_{\text{wazir}} = 0.3\), \(w_{\text{elephant}} = 0.25\), \(w_{\text{knight}} = 0.1\), \(w_{\text{camel}} = 0.2\), \(w_{\text{horse}} = 0.15\). Weight decay: \(g_t \gets g_t + \alpha \theta_t\).

\subsection{NIRMAL Optimizer}

\textbf{NIRMAL Optimizer} represents the original version, whose update rules are identical to the enhanced variant, except for the absence of damping factor.  
\textbf{Update Rules:}
\begin{align}
m_t &= \mu m_{t-1} + (1 - \mu) g_t, \\
v_t &= \beta v_{t-1} + (1 - \beta) g_t^2, \\
\Delta_{\text{wazir}} &= -\eta g_t, \\
\Delta_{\text{elephant}} &= -\eta m_t, \\
\Delta_{\text{knight}} &= \eta \kappa \mathcal{N}(0, 1), \\
\Delta_{\text{camel}} &= -\eta \gamma \frac{m_t}{\sqrt{v_t + \epsilon}}, \\
\Delta_{\text{horse}} &= -\eta \lambda \tanh(m_t), \\
\Delta_{\text{total}} &= w_{\text{wazir}} \Delta_{\text{wazir}} + w_{\text{elephant}} \Delta_{\text{elephant}} + w_{\text{knight}} \Delta_{\text{knight}} + w_{\text{camel}} \Delta_{\text{camel}} + w_{\text{horse}} \Delta_{\text{horse}}, \\
\theta_{t+1} &= \theta_t + \Delta_{\text{total}}.
\end{align}
Hyperparameters match Enhanced NIRMAL, excluding damping terms.

\subsection{Adam Optimizer}
\textbf{Adam (Adaptive Moment Estimation)} computes adaptive learning rates for each parameter by maintaining exponentially decaying moving averages of the past gradients (\(m_t\)) and the past squared gradients (\(v_t\)). Bias correction is applied to these moving averages to account for their initialization at zero. The update rule is given by:
\begin{align}
    m_t &= \beta_1 m_{t-1} + (1 - \beta_1) g_t, \label{eq:adam_momentum} \\
    v_t &= \beta_2 v_{t-1} + (1 - \beta_2) g_t^2, \label{eq:adam_variance} \\
    \hat{m}_t &= \frac{m_t}{1 - \beta_1^t}, \quad \hat{v}_t = \frac{v_t}{1 - \beta_2^t}, \label{eq:adam_bias_correction} \\
    \theta_{t+1} &= \theta_t - \eta \frac{\hat{m}_t}{\sqrt{\hat{v}_t} + \epsilon}. \label{eq:adam_update}
\end{align}
Standard hyperparameters for Adam are used: learning rate \(\eta = 10^{-3}\), decay rates for moment estimates \(\beta_1 = 0.9\) and \(\beta_2 = 0.999\), and a small constant \(\epsilon = 10^{-8}\).

\subsection{SGD with Momentum}
\textbf{Stochastic Gradient Descent (SGD) with Momentum} is a traditional but effective optimization algorithm. It aims to speed up the SGD in the relevant direction and dampen oscillations. This is achieved by adding a fraction of the update vector of the previous time step to the current update. The velocity term accumulates gradients over time, providing inertia \citep{soudani2019image}. Its update rules are as follows:
\begin{align}
    v_t &= \mu v_{t-1} + \eta g_t, \label{eq:sgd_momentum} \\
    \theta_{t+1} &= \theta_t - v_t. \label{eq:sgd_update}
\end{align}
The parameters used are: learning rate \(\eta = 0.01\) and momentum factor \(\mu = 0.9\). For the CIFAR datasets, a weight decay of \(\alpha = 5 \times 10^{-4}\) was applied to prevent overfitting.

\subsection{Nesterov Optimizer}
\textbf{Nesterov} uses a look-ahead approach. It is an updated version of momentum-based optimization.
\begin{align}
v_t &= \mu v_{t-1} + \eta g_t(\theta_t - \mu v_{t-1}), \\
\theta_{t+1} &= \theta_t - v_t.
\end{align}
Hyperparameters: \(\eta = 10^{-3}\), \(\mu = 0.9\), \(\alpha = 5 \times 10^{-4}\) (CIFAR datasets).

\section{Experimental Setup}
Our experiments were systematically conducted on four widely used image classification datasets to ensure a comprehensive evaluation.
\begin{itemize}
    \item \textbf{MNIST}: 70,000 grayscale digit images (60,000 train, 10,000 test), 10 classes.
    \item \textbf{FashionMNIST}: 70,000 grayscale fashion images (60,000 train, 10,000 test), 10 classes.
    \item \textbf{CIFAR-10}: 60,000 32x32 color images (50,000 train, 10,000 test), 10 classes.
    \item \textbf{CIFAR-100}: 60,000 32x32 color images (50,000 train, 10,000 test), 100 classes.
\end{itemize}

\subsection{Convolutional Neural Network (CNN) Architectures}
Custom CNNs were designed:
\begin{itemize}
    \item \textbf{MNIST/FashionMNIST}: Two convolutional layers (32, 64 filters), max pooling, two fully connected layers (128, 10 units).
    \item \textbf{CIFAR-10}: Three convolutional layers (64, 128, 256 filters), batch normalization, max pooling, two fully connected layers (512, 10 units).
    \item \textbf{CIFAR-100}: Four convolutional layers (64, 128, 256, 512 filters), batch normalization, max pooling, three fully connected layers (1024, 512, 100 units).
\end{itemize}

\subsection{Training Parameters}
All models were trained for 10 epochs using a batch size of 64. The cross-entropy loss function was utilized for training. To improve generalization and prevent overfitting, data augmentation techniques (random cropping and horizontal flipping) were applied specifically to the CIFAR-10 and CIFAR-100 datasets.

\subsection{Evaluation Metrics}
Performance was evaluated using the following:
\begin{itemize}
    \item \textbf{Test Accuracy}: Percentage of correctly classified images.
    \item \textbf{Test Loss}: Loss function value on the test set.
    \item \textbf{Weighted F1-score}: Harmonic mean of precision and recall, weighted by class size.
\end{itemize}

\subsection{Optimizer Hyperparameters}
Enhanced NIRMAL: \(\eta = 10^{-3}\), \(\mu = 0.9\), \(\beta = 0.999\), \(\epsilon = 10^{-8}\), \(\kappa = 0.01\), \(\gamma = 1.5\), \(\lambda = 0.5\), \(\alpha = 0.5\), \(r = 2.0\), weights \(w_{\text{wazir}} = 0.3\), \(w_{\text{elephant}} = 0.25\), \(w_{\text{knight}} = 0.1\), \(w_{\text{camel}} = 0.2\), \(w_{\text{horse}} = 0.15\), weight decay \(\alpha = 5 \times 10^{-4}\) (CIFAR datasets). NIRMAL: same as Enhanced NIRMAL, excluding damping terms. Adam: \(\eta = 10^{-3}\), \(\beta_1 = 0.9\), \(\beta_2 = 0.999\), \(\epsilon = 10^{-8}\). SGD with Momentum: \(\eta = 0.01\), \(\mu = 0.9\), \(\alpha = 5 \times 10^{-4}\) (CIFAR datasets). Nesterov: \(\eta = 10^{-3}\), \(\mu = 0.9\), \(\alpha = 5 \times 10^{-4}\) (CIFAR datasets).

\section{Results and Analysis}

This section presents and analyzes the experimental results, comparing the performance of Enhanced NIRMAL, NIRMAL, Adam, SGD with Momentum and Nestrov across the four benchmark datasets. Figure 1 shows test accuracy trends for optimizers in the MNIST, FashionMNIST, CIFAR-10, and CIFAR-100 datasets, and Table 1 summarizes key performance metrics, in terms of test accuracy, test loss, and F1-score (weighted) followed by detailed analyzes for each dataset. A comparative metrics plot provides an overview of performance across all datasets.

\begin{figure}[h]
    \centering
    \begin{subfigure}[b]{0.40\textwidth}
        \centering
        \includegraphics[width=\textwidth]{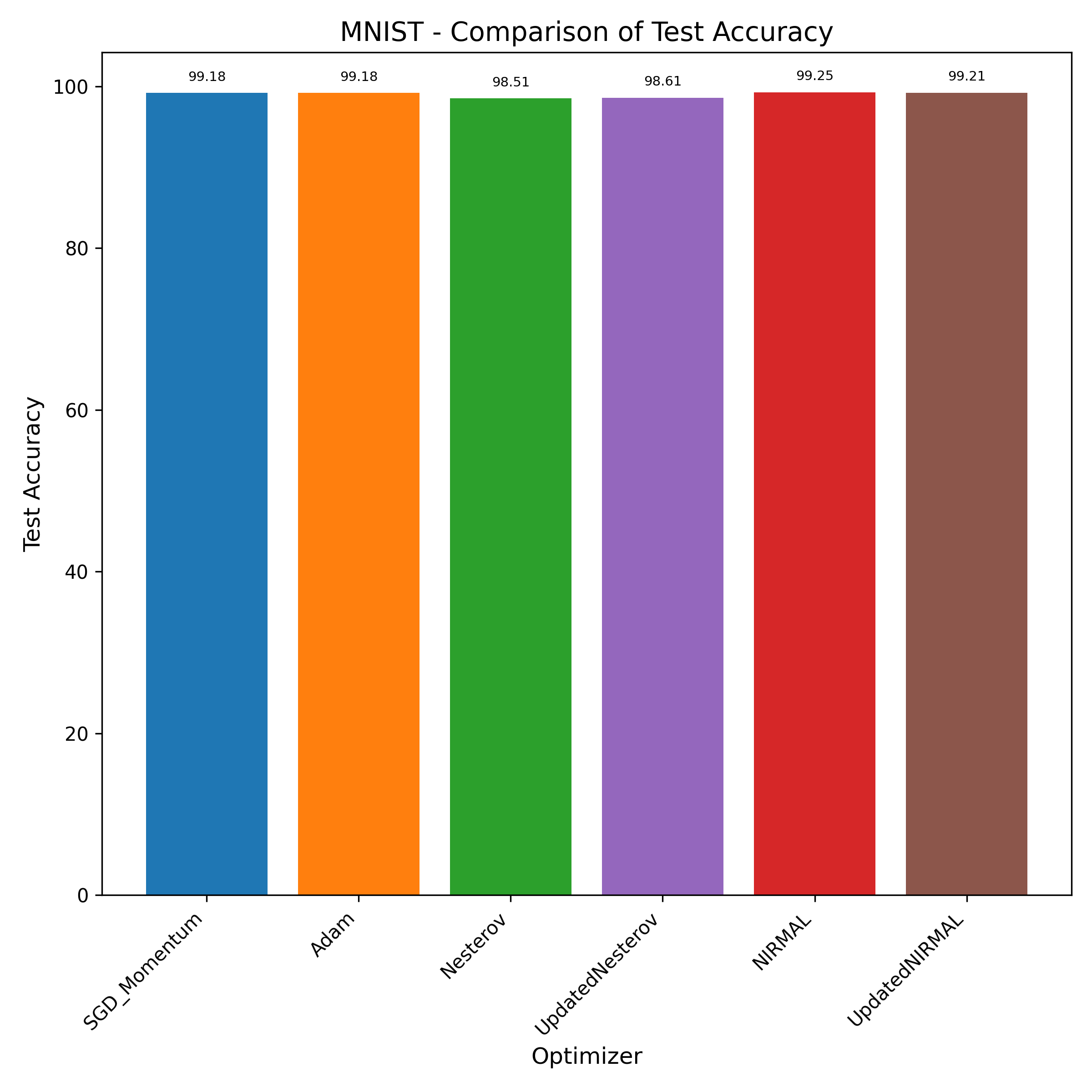}
        \caption{Test accuracy on MNIST}
        \label{fig:mnist_accuracy}
    \end{subfigure}
    \hfill
    \begin{subfigure}[b]{0.40\textwidth}
        \centering
        \includegraphics[width=\textwidth]{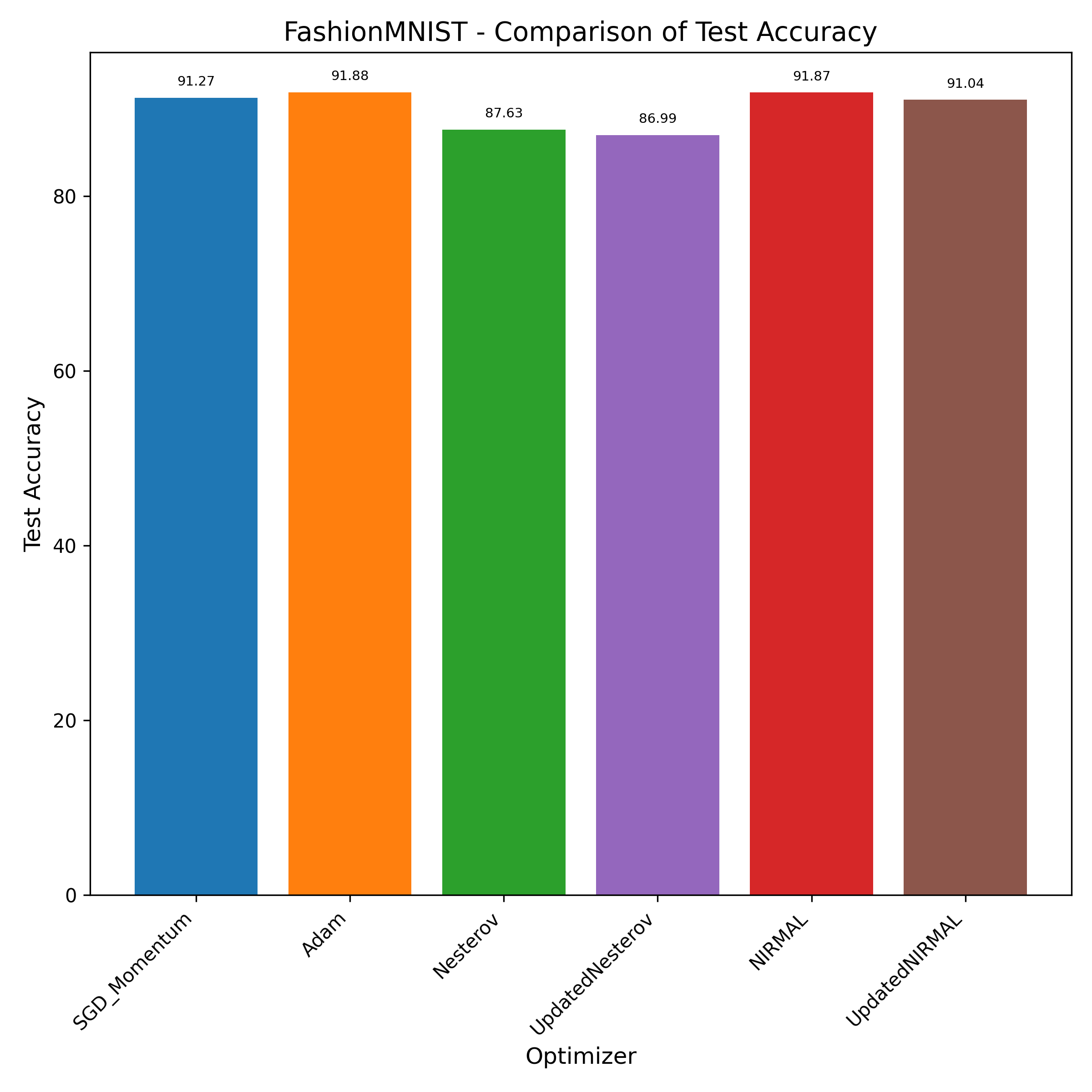}
        \caption{Test accuracy on FashionMNIST}
        \label{fig:fashionmnist_accuracy}
    \end{subfigure}
    
    \vspace{0.5cm}
    
    \begin{subfigure}[b]{0.40\textwidth}
        \centering
        \includegraphics[width=\textwidth]{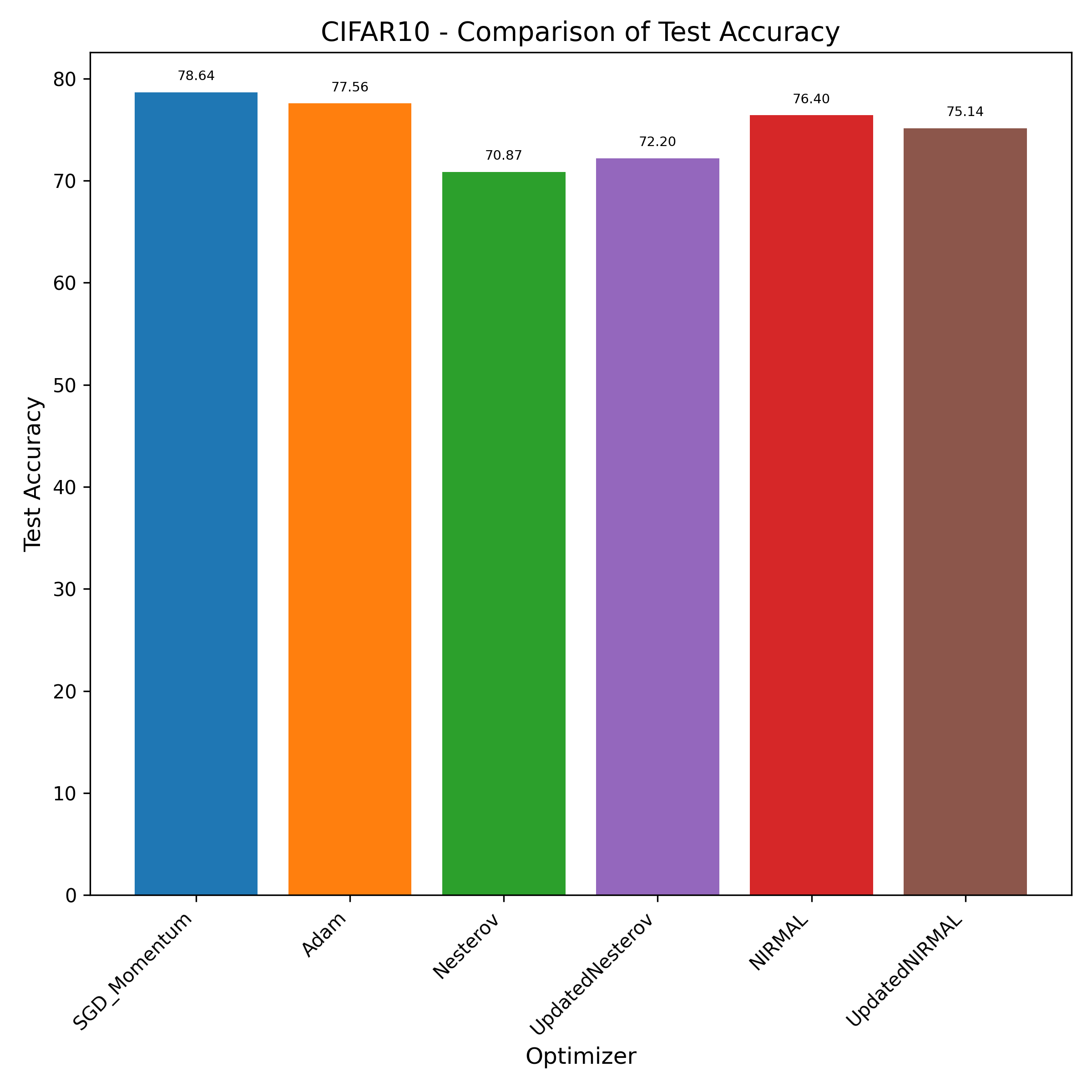}
        \caption{Test accuracy on CIFAR-10}
        \label{fig:cifar10_accuracy}
    \end{subfigure}
    \hfill
    \begin{subfigure}[b]{0.40\textwidth}
        \centering
        \includegraphics[width=\textwidth]{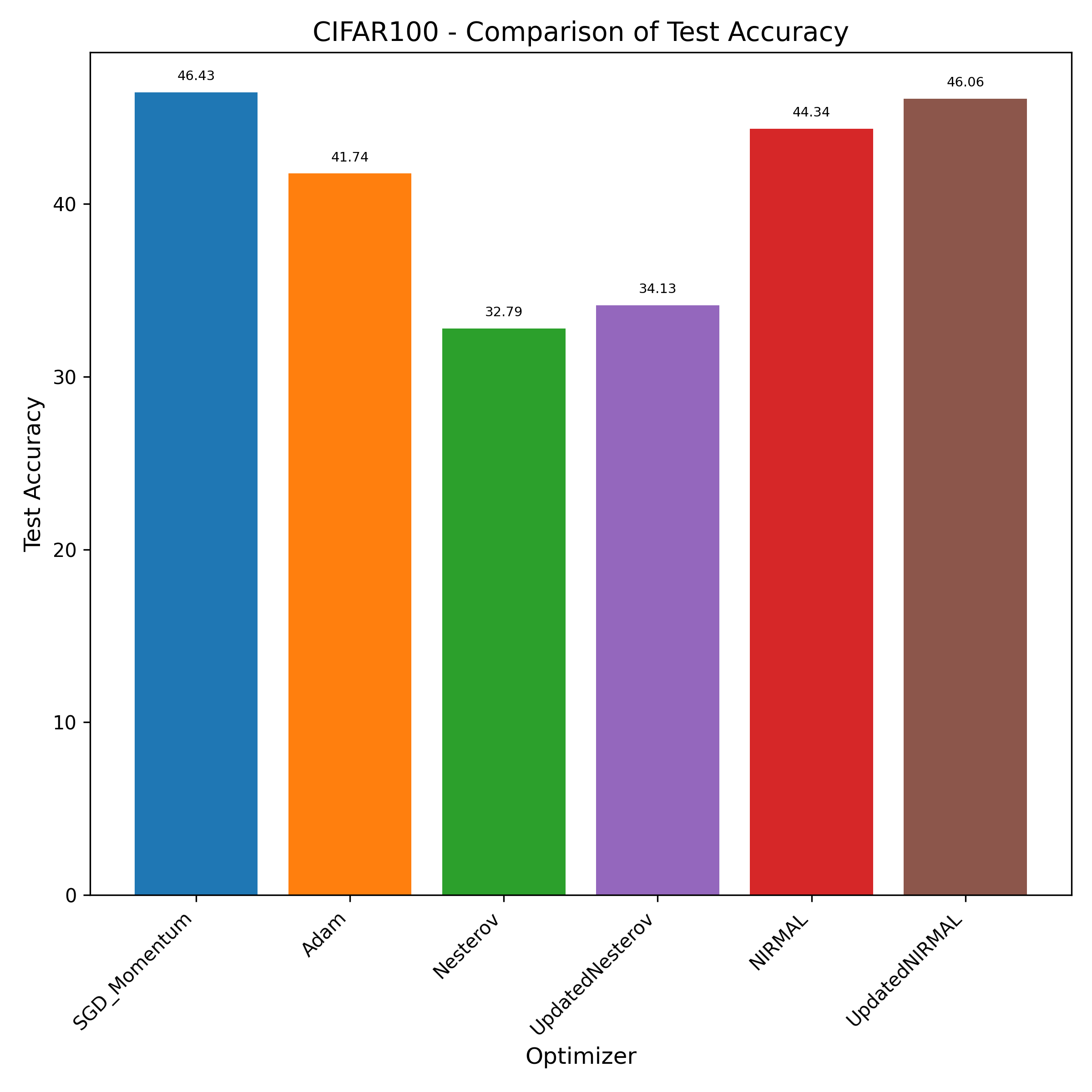}
        \caption{Test accuracy on CIFAR-100}
        \label{fig:cifar100_accuracy}
    \end{subfigure}
    \caption{Test accuracy trends for optimizers across MNIST, FashionMNIST, CIFAR-10, and CIFAR-100 datasets.}
    \label{fig:test_accuracy}
\end{figure}

\begin{table}[ht]
    \centering
    \caption{Comparative Performance of Optimizers Across Datasets}
    \label{tab:results}
    \begin{tabular}{lcccc}
        \toprule
        Dataset & Optimizer & Test Accuracy (\%) & Test Loss & F1-Score (weighted) \\
        \midrule
        \multirow{5}{*}{MNIST} 
        & SGD with Momentum & 99.18 & 0.024287 & 0.991800 \\
        & Adam & 99.18 & 0.032398 & 0.991802 \\
        & Nesterov & 98.51 & 0.040681 & 0.985103 \\
        & NIRMAL & 99.25 & 0.022567 & 0.992501 \\
        & Enhanced NIRMAL & 99.21 & 0.023685 & 0.992096 \\
        \midrule
        \multirow{5}{*}{FashionMNIST} 
        & SGD with Momentum & 91.27 & 0.255584 & 0.912266 \\
        & Adam & 91.88 & 0.257271 & 0.918006 \\
        & Nesterov & 87.63 & 0.334948 & 0.873894 \\
        & NIRMAL & 91.87 & 0.230960 & 0.918473 \\
        & Enhanced NIRMAL & 91.04 & 0.247975 & 0.910514 \\
        \midrule
        \multirow{5}{*}{CIFAR-10} 
        & SGD with Momentum & 78.64 & 0.616621 & 0.785790 \\
        & Adam & 77.56 & 0.647967 & 0.774931 \\
        & Nesterov & 70.87 & 0.800481 & 0.708163 \\
        & NIRMAL & 76.40 & 0.671712 & 0.765303 \\
        & Enhanced NIRMAL & 75.14 & 0.717061 & 0.748730 \\
        \midrule
        \multirow{5}{*}{CIFAR-100} 
        & SGD with Momentum & 46.43 & 1.979593 & 0.454754 \\
        & Adam & 41.74 & 2.215378 & 0.404318 \\
        & Nesterov & 32.79 & 2.590641 & 0.318412 \\
        & NIRMAL & 44.34 & 2.058203 & 0.432715 \\
        & Enhanced NIRMAL & 46.06 & 1.960435 & 0.442779 \\
        \bottomrule
    \end{tabular}
\end{table}

NIRMAL excels on MNIST (99.25\% accuracy, 0.022567 loss, 0.992501 F1-score) and FashionMNIST (91.87\% accuracy, 0.230960 loss, 0.918473 F1-score), slightly outperforming Enhanced NIRMAL (99.21\%, 0.023685, 0.992096 on MNIST; 91.04\%, 0.247975, 0.910514 on FashionMNIST) and Adam, while SGD with Momentum remains competitive and Nesterov lags significantly (98.51\%, 0.040681, 0.985103 on MNIST). On CIFAR-10, SGD with Momentum leads (78.64\% accuracy, 0.616621 loss, 0.785790 F1-score), followed by Adam and NIRMAL, with Enhanced NIRMAL (75.14\%, 0.717061, 0.748730) and Nesterov trailing. For CIFAR-100, Enhanced NIRMAL achieves the lowest loss (1.960435) and competitive accuracy (46.06\%) and F1-score (0.442779), nearly matching SGD with Momentum (46.43\%, 1.979593, 0.454754), while NIRMAL, Adam, and Nesterov underperform. These results highlight the superior stability of Enhanced NIRMAL’s in complex datasets such as CIFAR-100 due to its damped Nesterov acceleration, while NIRMAL and Adam perform well in simpler tasks, and SGD with Momentum is robust across complex datasets. Nesterov consistently struggles, particularly on CIFAR datasets. 

\section{Conclusion}

This study presents Enhanced NIRMAL, an advanced version of the NIRMAL optimizer \citep{gaud2025nirmal}. The primary improvement is the integration of an $(\alpha, r)$-damped Nesterov acceleration mechanism, inspired by \citep{cheng2025nesterov} and the Regularised Update Descent framework \citep{botev2016rud}. This damping reduces oscillations, improving stability during training.

The results highlight Enhanced NIRMAL's superior stability and generalization on complex datasets like CIFAR-100, attributed to its damped Nesterov mechanism. On simpler datasets (MNIST, FashionMNIST), the damping introduces slight overhead, making original NIRMAL and Adam viable alternatives. SGD with Momentum showed robustness on CIFAR datasets, while Nesterov struggled with complex tasks.

Future work includes optimizing hyperparameters (\(\alpha\), \(r\), and component weights) using grid search or Bayesian methods to further enhance performance, extending evaluations to larger and more diverse datasets such as ImageNet, and exploring adaptive damping schedules to balance stability and speed in varying dataset complexities. Additionally, integrating insights from recent optimization frameworks, such as those in \citep{botev2016rud}, could refine the damping mechanism, potentially incorporating second-order information or dynamic regularization to further improve convergence on non-convex problems.

\bibliographystyle{plainnat}
\bibliography{references}

\end{document}